\documentclass[a4paper,11pt]{article}
\pdfoutput=1

\usepackage{jheppub}

\usepackage{amsmath,amssymb,amsfonts}
\usepackage{graphicx}
\usepackage{booktabs}
\usepackage[colorlinks=true,linkcolor=blue,citecolor=blue,urlcolor=blue]{hyperref}

\title{Joint reconstruction of $H(z)$ and $f\sigma_8(z)$ with
physics informed neural networks}

\author{Konstantinos F. Dialektopoulos}
\affiliation{Institute of Space Sciences and Astronomy, University of Malta, Malta, MSD 2080}
\emailAdd{kdial01@um.edu.mt}

\abstract{%
We present a model-independent joint reconstruction of the Hubble parameter $H(z)$ and the growth rate $f\sigma_8(z)$ using a dual-head physics-informed neural network (PINN) trained on four complementary late-universe datasets: Cosmic Chronometers, Redshift-Space Distortions, the DESI DR2 BAO mean vector and full covariance, and the Pantheon$+$SH0ES supernova compilation. The two output heads share a backbone and are coupled through the linear growth equation of general relativity, penalizing the ODE residual at 1000 collocation points per training step via automatic differentiation. Uncertainty is quantified by an ensemble of 100 networks, each trained on an independent parametric-bootstrap resample of the data and its own draw of the fiducial cosmological parameters from Planck 2018 priors, so that the ensemble spread captures data-noise, initialization, and fiducial-cosmology systematics simultaneously. The physics coupling weight $\lambda$ is selected via an L-curve analysis over six values; the curve is nearly flat in total data $\chi^2$ (0.5\% variation across the full range), indicating that the joint dataset is intrinsically consistent with the growth ODE. We adopt $\lambda = 0.1$, which reduces the ODE residual by 85\% at a data-$\chi^2$ cost of 0.2\%. Without any $H_0$ prior, the free reconstruction yields $H_0 = 69.0 \pm 4.7$\,km\,s$^{-1}$\,Mpc$^{-1}$ (ensemble median 70.7\,km\,s$^{-1}$\,Mpc$^{-1}$), consistent with the Planck 2018 CMB value and with the DESI DR2 inverse distance-ladder determination, and approximately $0.9\sigma$ below the SH0ES local measurement. The reconstructed $H(z)$ lies systematically below the flat $\Lambda$CDM prediction at $z \sim 0.7$--$0.8$ ($\approx 1.6\sigma$), consistent with the dark energy evolution suggested by DESI DR2. As conditional analyses, we also anchor $H_0$ to the SH0ES value $73.04 \pm 1.04$\,km\,s$^{-1}$\,Mpc$^{-1}$ and the Local Distance Network consensus $73.50 \pm 0.81$\,km\,s$^{-1}$\,Mpc$^{-1}$; both anchored reconstructions yield a suppressed $f\sigma_8$, illustrating the propagation of the $H_0$--$\sigma_8$ link through the ODE coupling.
}

\begin{document}
\maketitle

\section{Introduction}
\label{sec:intro}

The $\Lambda$CDM concordance model describes an impressive range of observations with only six parameters. Yet two persistent discrepancies between early- and late-universe measurements have accumulated enough statistical weight to be taken seriously as potential signs of new physics. The first is the $H_0$ tension: local distance-ladder measurements return $H_0 \approx 73$\,km\,s$^{-1}$\,Mpc$^{-1}$; the SH0ES team reports $73.04 \pm 1.04$\,km\,s$^{-1}$\,Mpc$^{-1}$ \cite{Riess:2021jrx} and the Local Distance Network finds $73.50 \pm 0.81$\,km\,s$^{-1}$\,Mpc$^{-1}$ \cite{H0DN:2025lyy}, while Planck CMB analyses under $\Lambda$CDM yield $67.4 \pm 0.5$\,km\,s$^{-1}$\,Mpc$^{-1}$ \cite{Planck:2018vyg}, a disagreement exceeding $5\sigma$. The second is the $\sigma_8$ tension: galaxy weak-lensing surveys find a lower amplitude of matter fluctuations than Planck predicts \cite{Heymans:2020gsg, Abbott:2021qlj}, pointing to less growth at late times than $\Lambda$CDM expects. Comprehensive reviews of these tensions are given in Refs.~\cite{CosmoVerseNetwork:2025alb, Perivolaropoulos:2021jda}.

A third piece of evidence has recently been added by the Dark Energy Spectroscopic Instrument. The DESI DR2 BAO measurements \cite{DESI:2025zgx}, when combined with CMB and supernova data, show a preference for dynamical dark energy ($w \neq -1$) at the $2$--$3\sigma$ level, suggesting that the expansion history departs from flat $\Lambda$CDM at intermediate redshifts.

A natural response to all three anomalies is to reconstruct the relevant observables directly from late-universe data, without assuming a specific dark energy model, and check whether the result is consistent with $\Lambda$CDM. The Hubble parameter $H(z)$ encodes the expansion history; the combination $f\sigma_8(z)$ encodes the linear growth rate as measured by redshift-space distortions. Reconstructing both functions simultaneously and probing their consistency with the governing equations is one of the cleanest available diagnostics.

Gaussian processes have long been the tool of choice for such reconstructions \cite{Seikel:2012uu}, and artificial neural networks (ANNs) have more recently provided flexible, uncertainty-aware alternatives. In \cite{Dialektopoulos:2021wde} the authors performed a careful ANN reconstruction of both $H(z)$ and $f\sigma_8(z)$, applied the $\mathrm{Om}(z)$ and $\mathrm{Om}_{f\sigma_8}(z)$ null tests to the resulting ensembles, and found significant evidence for a departure from $\Lambda$CDM in the growth sector.

What neither Gaussian processes nor conventional ANNs exploit is the known physical coupling between the two observables. In general relativity, $H(z)$ and $f\sigma_8(z)$ are linked through the linear growth equation \cite{Linder:2005in},
\begin{equation}  \label{eq:growth_ode}
  \frac{{\rm d}f}{{\rm d}\ln a} + f^2 + \left[2 + \frac{1}{2}\frac{{\rm d}\ln H^2}{{\rm d}\ln a}\right]f = \frac{3}{2}\,\Omega_m(a)\,,
\end{equation}
where $\Omega_m(a) = \Omega_{m,0}\,(1+z)^3\,H_0^2/H^2(z)$ and $f \equiv {\rm d}\ln\delta_m/{\rm d}\ln a$. Any information the $H(z)$ data carry about the growth history through this equation is discarded in an independent reconstruction.

Physics-informed neural networks (PINNs) \cite{Raissi:2019, Lagaris:1998} offer a remedy. By adding the ODE residual to the training loss, one couples the two output heads of a shared network through the growth equation directly during optimization, rather than as a post-hoc consistency check. The coupling strength is controlled by a scalar weight $\lambda$, chosen here via a systematic L-curve analysis.

In this paper we apply this dual-head PINN to the most constraining currently available late-universe datasets: DESI DR2 BAO with full covariance, the Pantheon$+$SH0ES supernova compilation with full STAT$+$SYS covariance, cosmic chronometer $H(z)$ measurements, and a compilation of 22 RSD $f\sigma_8$ measurements. We train an ensemble of 100 networks with parametric-bootstrap data resampling and per-member fiducial-cosmology draws to propagate the dominant systematics into the uncertainty bands.

The main results are: (i) a free (no-prior) reconstruction yields $H_0 = 69.0 \pm 4.7$ km s$^{-1}$ Mpc$^{-1}$ from the inverse distance ladder, consistent with Planck and DESI, and approximately $0.9\sigma$ below SH0ES; (ii) the reconstructed $H(z)$ lies approximately $1.6\sigma$ below the flat $\Lambda$CDM prediction at $z \sim 0.7$--$0.8$, consistent with the dark energy evolution reported by DESI DR2; (iii) the L-curve analysis confirms that the joint dataset is intrinsically consistent with the growth ODE, with $\lambda = 0.1$ providing the optimal data-fit/physics-consistency balance.

Physics-informed neural networks have rapidly gained traction in cosmology. The approach was established in the cosmological context by Chantada et al.~\cite{Chantada:2022bdf}, and has since been extended to Bayesian parameter inference \cite{Chantada:2023hhh}, dark-energy equation-of-state reconstruction from DESI and supernova data \cite{Paliathanasis:2026dqk, Verma:2025ujt, Tan:2025xas}, and supernova absolute magnitude reconstruction \cite{Staicova:2026gwv}. Beyond background reconstruction, PINNs have been applied to dark-matter phenomenology \cite{Bento:2025agw}, fuzzy dark matter collapse \cite{Mishra:2025ipo}, Vlasov--Poisson evolution \cite{Cerardi:2025ato}, and baryonic content inference \cite{Dai:2023kip}. Pure data-driven neural reconstructions of late-time cosmology from Pantheon data \cite{Dialektopoulos:2023dhb}, scalar-tensor gravity \cite{Dialektopoulos:2023jam}, the Hubble rate and its derivative \cite{Mukherjee:2022yyq}, and a possible late-time $M_B$ transition \cite{Mukherjee:2024akt} provide the comparison baseline.  No prior PINN reconstruction has jointly modeled the expansion history and the linear growth of structure under a shared physical constraint; this is the gap the present work addresses.

The paper is organized as follows. Section~\ref{sec:pinns} gives a brief overview of PINNs. Section~\ref{sec:method} describes the architecture, data, loss functions, and training procedure. Section~\ref{sec:results} presents the results. Section~\ref{sec:discussion} discusses the findings and outlines next steps. Null tests are collected in Appendix~\ref{app:nulltests}. Throughout, $H(z)$ is in km\,s$^{-1}$\,Mpc$^{-1}$.

\section{Physics Informed Neural Networks}
\label{sec:pinns}

The idea of using neural networks to solve differential equations goes back at least to \cite{Lagaris:1998}, who showed that a network can satisfy an ODE or PDE by incorporating the equation residual into its training objective. The framework was formalized and named by Raissi, Perdikaris \& Karniadakis \cite{Raissi:2019}.

The core idea is as follows. Let $u_\theta$ be a neural network approximating the solution of a differential equation $\mathcal{L}[u] = 0$, and suppose one also has observational data $\{(x_i,y_i)\}$. A PINN minimizes
\begin{equation}  \label{eq:pinn_loss_general}
  \mathcal{L}_{\rm total} = \mathcal{L}_{\rm data} + \lambda\,\mathcal{L}_{\rm physics}\,,
\end{equation}
where the physics loss is the mean squared ODE residual at $N_c$ collocation points:
\begin{equation}  \label{eq:phys_loss_general}
  \mathcal{L}_{\rm physics} = \frac{1}{N_c}\sum_{i=1}^{N_c}
  \bigl|\mathcal{L}[u_\theta](x_i^c)\bigr|^2.
\end{equation}
Evaluating $\mathcal{L}_{\rm physics}$ requires the derivatives ${\rm d}f/{\rm d}\ln a$ and ${\rm d}\ln H^2/{\rm d}\ln a$ at each collocation point. Because the network is a composition of smooth analytic functions, automatic differentiation computes these via the chain rule exactly through the computation graph, with no approximation beyond floating-point arithmetic and no spatial mesh. The ELU activation used here is infinitely differentiable, so the ODE residual is well-defined at every collocation point.

The weight $\lambda$ controls the trade-off between data fit and physics consistency. At $\lambda = 0$ the network is a standard unconstrained ANN; as $\lambda$ grows, it is progressively forced toward ODE-satisfying solutions. The appropriate $\lambda$ is determined here via an L-curve analysis rather than asserted \textit{a priori} (Section~\ref{sec:ablation}).

The appeal for cosmological inverse problems is straightforward. GR and perturbation theory provide governing equations; the observational data are sparse and noisy; and one wants reconstructions consistent with both. The ODE acts as a physically motivated regularizer grounded in known physics. If the physics turns out not to hold exactly, which is what one may wish to test, the data term in Eq.~\eqref{eq:pinn_loss_general} will push back, and the tension between data and ODE will be visible in the loss balance.

\section{Methodology}
\label{sec:method}

This section describes the full pipeline. Section~\ref{sec:data} presents the   four observational data blocks used in training. Section~\ref{sec:arch} details the network architecture. Section~\ref{sec:physics_loss} derives the physics loss from the linear growth ODE and explains how the required derivatives are computed via automatic differentiation. Section~\ref{sec:data_loss} defines the four data losses and the total training objective. Section~\ref{sec:training} describes the optimization procedure and the bootstrap and prior propagated ensemble used for uncertainty quantification. The pipeline is implemented in PyTorch; all derivatives needed by the physics loss and the distance integral are computed through the network's own computation graph, with no numerical differentiation or spatial mesh required.

\subsection{Data}
\label{sec:data}

\paragraph{Cosmic chronometers.}
The $H(z)$ compilation contains 32 cosmic chronometer (CC) measurements obtained by applying $H(z) = -(1+z)^{-1}\,dz/dt$ to the differential ages of passively evolving galaxies \cite{Jimenez:2001gg, Moresco:2012jh, Moresco:2015cya, Moresco:2016mzx, Moresco:2020fbm, Borghi:2021rft}, spanning $z \in [0.07,\,1.97]$. These measurements are purely kinematic and carry no sound-horizon calibration.

\paragraph{DESI DR2 BAO.}
We include the DESI DR2 BAO mean vector and full covariance matrix \cite{DESI:2025zgx}. The 13-element mean vector contains distance ratios at six effective redshifts: $D_V/r_s$ at $z_{\rm eff} = 0.295$ (BGS); $D_M/r_s$ and $D_H/r_s$ at $z_{\rm eff} = 0.510$ (LRG1), $0.706$ (LRG2), $0.934$ (LRG3$+$ELG1), $1.321$ (ELG2), and $1.484$ (QSO); and $D_H/r_s$ and $D_M/r_s$ at $z_{\rm eff} = 2.330$ (Ly$\alpha$). Here $D_H(z) = c/H(z)$ and $D_M(z) = (1+z)^{-1}\,D_L(z)$ is the comoving angular diameter distance, computed via the differentiable integral 
\begin{equation}
  D_M(z) = c\int_0^z \frac{{\rm d}z'}{H(z')}\,,
  \label{eq:dm}
\end{equation}
evaluated at $N_d = 3000$ quadrature points through the network's $H(z)$ head at every training step. This is compared to the data as $D_M/r_s$ and $D_H/r_s$, using the fiducial sound horizon $r_s = r_d = 147.09$\,Mpc (Planck 2018), which is drawn per ensemble member from $\mathcal{N}(147.09,\,0.26^2)$\,Mpc (Section~\ref{sec:training}).

\paragraph{Pantheon$+$SH0ES.}
We include the Pantheon$+$SH0ES compilation of Type Ia supernova distance moduli with full STAT$+$SYS covariance \cite{Brout:2022vxf}, cut to the standard Hubble-flow sample of 1580 supernovae by excluding the Cepheid-host calibrators and supernovae with $z_{\rm HD} < 0.01$. Each distance modulus is compared to the model prediction
\begin{equation}
  \mu(z) = 5\log_{10}\!\left[\frac{(1+z)\,D_M(z)}{10\,\mathrm{pc}}\right],
\end{equation}
where $D_M(z)$ is computed through the same differentiable integral as for the BAO comparison. The absolute magnitude $M_B$ is marginalized analytically at each training step following the standard Pantheon$+$ convention.

\paragraph{Growth rate.}
For $f\sigma_8(z)$ we use the 22-point RSD compilation of \cite{Akarsu:2025ijk}, aggregating measurements from spectroscopic galaxy surveys spanning $z \in [0.02,\,1.94]$. No Alcock--Paczy\'{n}ski corrections are applied to the individual measurements. We instead bound the size of this systematic. Following the first-order AP correction of \cite{Kazantzidis:2018rnb}, $f\sigma_8^{\rm corr}(z) = q(z)\,f\sigma_8^{\rm obs}(z)$ with $q(z) = [H(z)D_A(z)]_{\rm Planck18} / [H(z)D_A(z)]_{\rm fid}$, and noting that $H_0$ cancels from the product $H(z)D_A(z)$ for flat $\Lambda$CDM so that $q(z)$ depends only on each survey's fiducial $\Omega_{m,0}$ (all in the range $0.266$--$0.310$, against our Planck 2018 reference value $0.3153$), we find that homogenizing every point to the Planck 2018 fiducial would shift $f\sigma_8(z)$ by at most $2.8\%$ (root-mean-square $1.0\%$ across the 22 points), or at most $0.13\,\sigma_i$ (root-mean-square $0.06\,\sigma_i$) relative to each point's quoted uncertainty. This is an order of magnitude below the measurement precision at every redshift, so we report the published, uncorrected values.

\subsection{Network architecture}
\label{sec:arch}

\begin{figure}[t]
    \centering
    \includegraphics[width=0.85\linewidth]{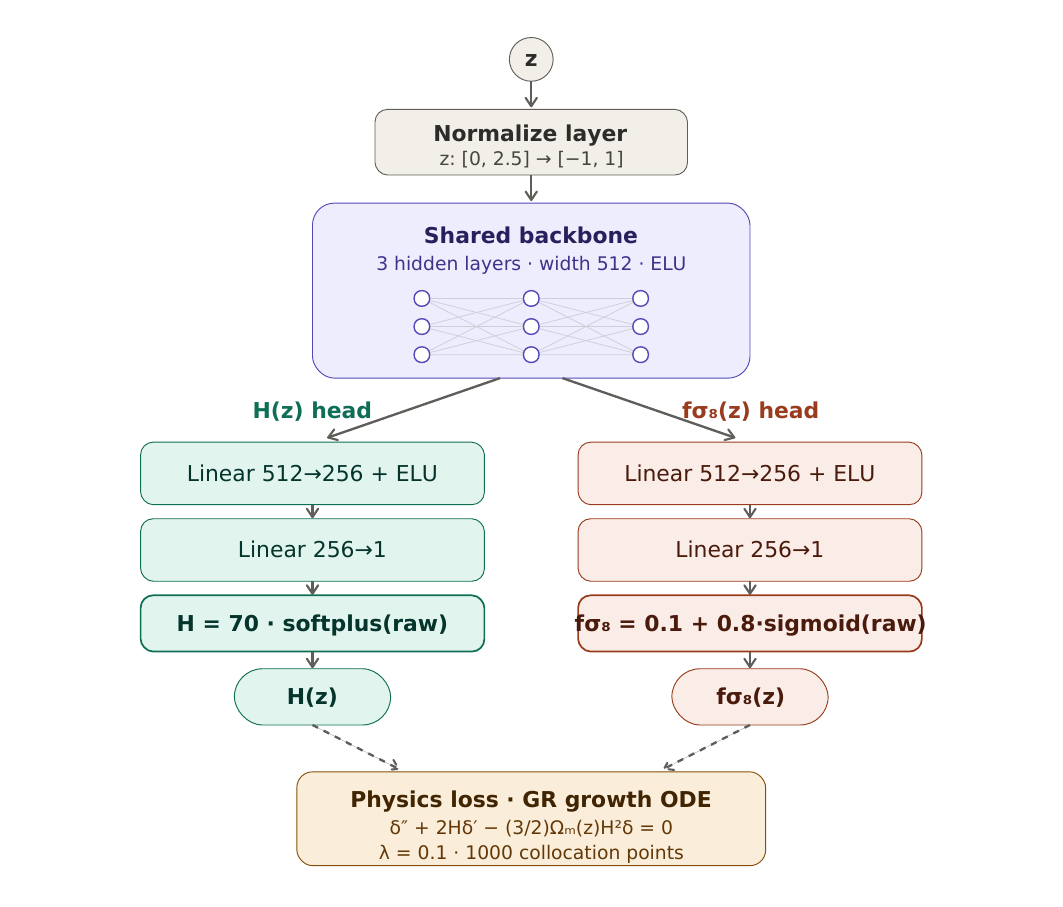}
    \caption{Architecture of the dual-head PINN. A shared backbone maps the input
    redshift to two task-specific heads reconstructing $H(z)$ and $f\sigma_8(z)$, with output activations enforcing positivity and physical bounds. The two heads are coupled through a physics loss that penalizes the residual of the GR linear growth equation.}
    \label{fig:pinn_arch}
\end{figure}

The network takes a single redshift $z$ as input and returns the pair $(H(z),\,f\sigma_8(z))$. A shared backbone of three fully connected layers of width 512, each followed by an ELU activation ($\alpha = 1$), feeds two independent output heads (each $\mathrm{Linear}(512\!\to\!256)$ with ELU followed by $\mathrm{Linear}(256\!\to\!1)$) that predict $H(z)$ and $f\sigma_8(z)$, respectively (Figure~\ref{fig:pinn_arch}). 

\begin{figure}[t]
  \centering
  \includegraphics[width=0.7\textwidth]{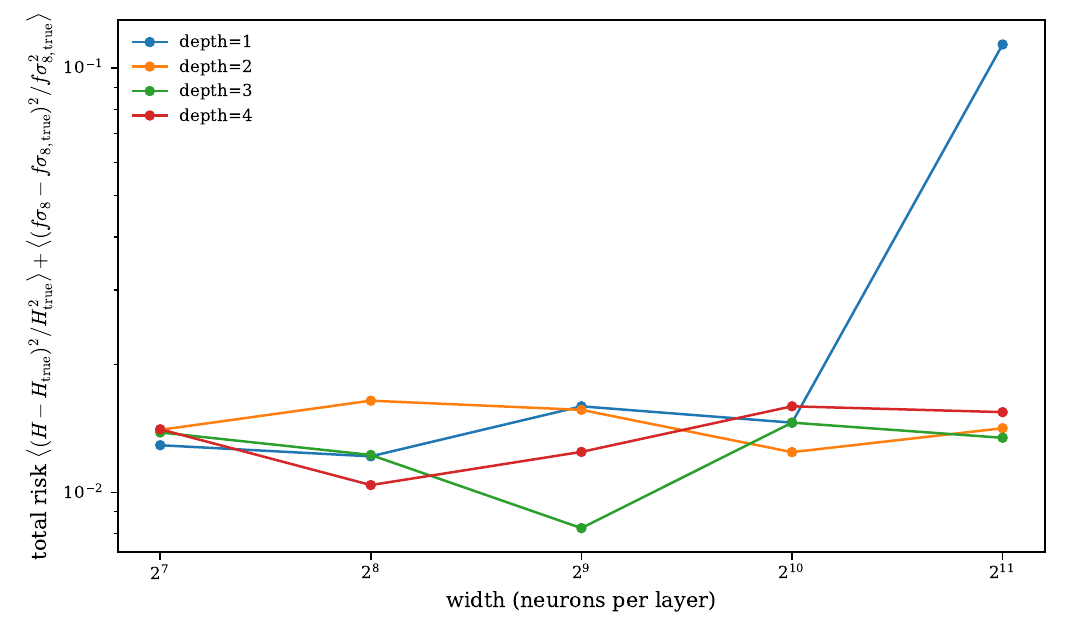}
  \caption{Recovery risk against the known mock ground truth as a function of
  backbone width, for each of the four depths tested, averaged over two mock fiducial cosmologies ($\Omega_{m,0} = 0.30$ and $0.35$). Risk is non-monotonic in both width and depth; the global minimum occurs at width $512$, depth $3$, the architecture adopted throughout this work.}
  \label{fig:arch_sweep}
\end{figure}

The choice of a three layer, width 512 backbone is not arbitrary. We validated it against mock data with a known ground truth: for two fiducial cosmologies ($\Omega_{m,0} = 0.30$ and $0.35$), we generated synthetic $H(z)$, $f\sigma_8(z)$, DESI DR2 BAO, and Pantheon$+$ compilations matching the real data in size and noise level, then trained small ensembles (3 seeds per cell, no bootstrap resampling, since here we want to isolate the architecture's own recovery error rather than propagate data uncertainty) across a grid of $20$ architectures spanning widths $\{128,\,256,\,512,\,1024,\,2048\}$ and depths $\{1,\,2,\,3,\,4\}$. For each cell we measured the mean relative squared error against the known mock truth, averaged over both fiducial cosmologies. The global minimum occurs at width $512$, depth $3$ (Figure~\ref{fig:arch_sweep}), with a total risk about $30$--$40\%$ lower than neighboring cells; risk does not decrease monotonically with either width or depth, so simply scaling up the network would not have improved recovery. We adopt this architecture, $\approx 8 \times 10^5$ parameters, for all production runs.

The input is affinely mapped to $\tilde{z} \in [-1,1]$ via
\begin{equation}
  \tilde{z} = \frac{2(z - z_{\rm min})}{z_{\rm max} - z_{\rm min}} - 1\,,
  \qquad z_{\rm min} = 0\,,\quad z_{\rm max} = 2.5\,.
  \label{eq:normalise}
\end{equation}
ELU activations are chosen because they are infinitely differentiable with non-zero gradients throughout the real line, which is essential for evaluating the ODE residual via automatic differentiation. The $H$ head output is scaled as
\begin{equation}
  H(z) = H_{\rm fid}\,\mathrm{softplus}(h_\theta(z))\,,
  \qquad H_{\rm fid} = 70\;\mathrm{km\,s^{-1}\,Mpc^{-1}}\,,
\end{equation}
guaranteeing strict positivity. The $f\sigma_8$ head is passed through a logistic function rescaled to the interval $(0.1,\,0.9)$:
\begin{equation}
  f\sigma_8(z) = 0.1 + 0.8\,\sigma(g_\theta(z))\,.
\end{equation}
The total parameter count is approximately $8 \times 10^5$, dominated by the three backbone layers.

\subsection{Physics loss}
\label{sec:physics_loss}

In flat GR with a smooth dark energy component, the linear growth rate $f$ satisfies Eq.~\eqref{eq:growth_ode}. Both logarithmic derivatives are computed via PyTorch automatic differentiation using the chain rule
\begin{equation}
  \frac{{\rm d}g}{{\rm d}\ln a} = -(1+z)\,\frac{{\rm d}g}{{\rm d}z}\,.
\end{equation}
At each training step, $N_c = 1000$ collocation points are sampled uniformly from $z \in [0,\,2.5]$, and the physics loss is 
\begin{equation}
  \mathcal{L}_{\rm physics} = \frac{1}{N_c}\sum_{i=1}^{N_c}\mathcal{R}^2(z_i)\,,
  \label{eq:phys_loss}
\end{equation}
where $\mathcal{R}(z)$ is the residual of Eq.~\eqref{eq:growth_ode} at the network outputs. The growth rate entering the ODE is obtained from the network's $f\sigma_8$ output via
\begin{equation}
  f(z) = \frac{f\sigma_8(z)}{\sigma_8(z)} = \frac{f\sigma_8(z)}{\sigma_{8,0}\,D(z)}\,,
\end{equation}
where $D(z)$ is the linear growth factor normalized to $D(0) = 1$, pre-computed from the $\Lambda$CDM background at the per-member fiducial cosmology. For the exact $\Lambda$CDM solution, $\mathcal{L}_{\rm physics} < 10^{-12}$.

The matter density $\Omega_m(a)$ uses the network's own $H(z)$, so the physics loss is model-independent in the dark energy equation of state; the only prior information entering the ODE is $\Omega_{m,0}$, $\sigma_{8,0}$, and $D(z)$, all of which are varied across ensemble members.

\subsection{Data losses and total training objective}
\label{sec:data_loss}

All four data blocks use a Gaussian $\chi^2$ loss. For the CC and RSD data, which have diagonal covariances, this reduces to
\begin{equation}
  \mathcal{L}_{\mathrm{data},H} = \sum_{i=1}^{N_H}
  \frac{\bigl[H(z_i) - H_{\rm obs,i}\bigr]^2}{\sigma_{H,i}^2}\,,
  \label{eq:loss_H}
\end{equation}
and analogously for $\mathcal{L}_{\mathrm{data},f\sigma_8}$.

For DESI DR2 BAO and Pantheon$+$SH0ES, the full covariance matrices $\mathbf{C}_{\rm DESI}$ and $\mathbf{C}_{\rm SN}$ are used:
\begin{align}
  \mathcal{L}_{\rm DESI} &= \boldsymbol{\Delta}_{\rm DESI}^{\top}
  \mathbf{C}_{\rm DESI}^{-1}\,\boldsymbol{\Delta}_{\rm DESI}\,, \label{eq:loss_desi}\\
  \mathcal{L}_{\rm SN}   &= \boldsymbol{\Delta}_{\rm SN}^{\top}
  \mathbf{C}_{\rm SN}^{-1}\,\boldsymbol{\Delta}_{\rm SN}\,,   \label{eq:loss_sn}
\end{align}
where $\boldsymbol{\Delta}$ denotes the vector of model-minus-data residuals. The DESI residuals are in $D_M/r_s$ and $D_H/r_s$, computed through the differentiable integral~\eqref{eq:dm}. The Pantheon$+$ residuals are in distance modulus $\mu(z)$, computed through the same integral after analytical marginalization over $M_B$.

The total training objective is
\begin{equation}
  \mathcal{L}_{\rm total} = \mathcal{L}_{\mathrm{data},H}
  + \mathcal{L}_{\mathrm{data},f\sigma_8}
  + \mathcal{L}_{\rm DESI}
  + \mathcal{L}_{\rm SN}
  + \lambda\,\mathcal{L}_{\rm physics}\,.
  \label{eq:total_loss}
\end{equation}
An $H_0$ prior, when used, is incorporated as an additional Gaussian term at $z = 0$ appended to $\mathcal{L}_{\mathrm{data},H}$, with central value and uncertainty set to the prior specification.

\subsection{Training and uncertainty quantification}
\label{sec:training}

All networks are optimized with Adam \cite{Kingma:2014vow} at an initial learning rate $\eta_0 = 10^{-3}$. A ReduceLROnPlateau scheduler halves the rate whenever $\mathcal{L}_{\rm total}$ fails to decrease for 500 consecutive epochs. Each network is trained for $5 \times 10^4$ epochs. The pipeline is implemented in PyTorch \cite{Paszke:2019xhk}; all derivatives are computed via automatic differentiation.

Uncertainty is quantified by an ensemble of 100 networks. Rather than varying only the initialization seed, each member $k$ receives:
\begin{enumerate}
\item A \emph{bootstrap resample} of the data: observed values are replaced by $y^* \sim \mathcal{N}(y_{\rm obs}, \mathbf{C})$, where $\mathbf{C}$ is the measurement covariance for each data block. This propagates data-noise uncertainty into the ensemble spread.
\item An \emph{independent draw of the fiducial cosmological parameters} from Planck 2018 priors: $\Omega_{m,0} \sim \mathcal{N}(0.3153,\,0.007^2)$, $\sigma_{8,0} \sim \mathcal{N}(0.8111,\,0.006^2)$, $r_d \sim \mathcal{N}(147.09,\,0.26^2)$\,Mpc. This propagates the fiducial-cosmology systematic into the ensemble spread, rather than treating it as a fixed assumption.
\end{enumerate}
The ensemble mean and standard deviation of $H(z)$ and $f\sigma_8(z)$ on a uniform grid of 500 redshifts in $[0,\,2.5]$ provide the central value and $1\sigma$ band. The resulting spread captures network-initialization variance, data-noise uncertainty, and the fiducial-cosmology systematic simultaneously. Key hyperparameters are collected in Table~\ref{tab:hyperparams}.

\begin{table}[t]
\centering
\caption{Network and training hyperparameters.}
\label{tab:hyperparams}
\begin{tabular}{lc}
\toprule
Hyperparameter & Value \\
\midrule
Backbone width & 512 \\
Backbone layers & 3 \\
Activation & ELU ($\alpha = 1$) \\
Fiducial $H_{\rm fid}$ / km\,s$^{-1}$\,Mpc$^{-1}$ & 70 \\
Fiducial $\Omega_{m,0}$ (central / $1\sigma$ prior) & $0.3153\;/\;0.007$ \\
Fiducial $\sigma_{8,0}$ (central / $1\sigma$ prior) & $0.8111\;/\;0.006$ \\
Fiducial $r_d$ / Mpc (central / $1\sigma$ prior) & $147.09\;/\;0.26$ \\
Collocation points $N_c$ & 1000 (resampled each step) \\
Distance quadrature points $N_d$ & 3000 \\
Initial learning rate & $10^{-3}$ \\
LR patience / epochs & 500 \\
LR reduction factor & 0.5 \\
Training epochs & $5 \times 10^4$ \\
Ensemble size & 100 \\
Bootstrap resample & yes \\
\bottomrule
\end{tabular}
\end{table}

\section{Results}
\label{sec:results}

We run six $\lambda$ configurations without an $H_0$ prior ($\lambda \in \{0,\,0.01,\,0.03,\,0.1,\,0.3,\,1.0\}$) to characterize the optimal physics coupling weight via an L-curve, and then two anchored runs at the selected $\lambda = 0.1$ with the SH0ES and H0dN priors as conditional analyses. The free reconstruction is the primary result. We begin with loss convergence, then present the L-curve and $\lambda$ selection, and finally the $H(z)$, $f\sigma_8(z)$, and ensemble results. Null tests are deferred to Appendix~\ref{app:nulltests}.

\subsection{Loss convergence}
\label{sec:loss_convergence}

\begin{figure}[t]
\centering
\includegraphics[width=0.85\textwidth]{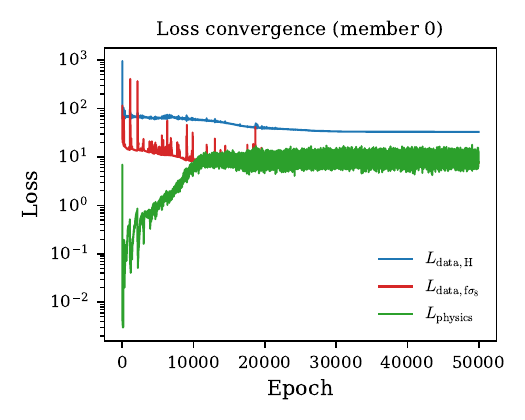}
\caption{Training loss history for ensemble member 0 at $\lambda = 0.1$ (no prior). All five components converge smoothly on a log scale over $5 \times 10^4$ epochs.}
\label{fig:loss}
\end{figure}

Figure~\ref{fig:loss} shows the five loss components for ensemble member 0 at $\lambda = 0.1$ (no prior). All components converge smoothly. The Pantheon$+$ $\chi^2$ term dominates by construction (1580 supernovae), while the CC, DESI, and $f\sigma_8$ terms converge to comparable magnitudes. The physics loss drops several orders of magnitude within the first $\sim 5000$ epochs and stabilizes. The learning rate reaches $\lesssim 10^{-7}$ at the final epoch, confirming convergence.

\subsection{L-curve analysis and $\lambda$ selection}
\label{sec:ablation}

\begin{figure}[t]
\centering
\includegraphics[width=0.60\textwidth]{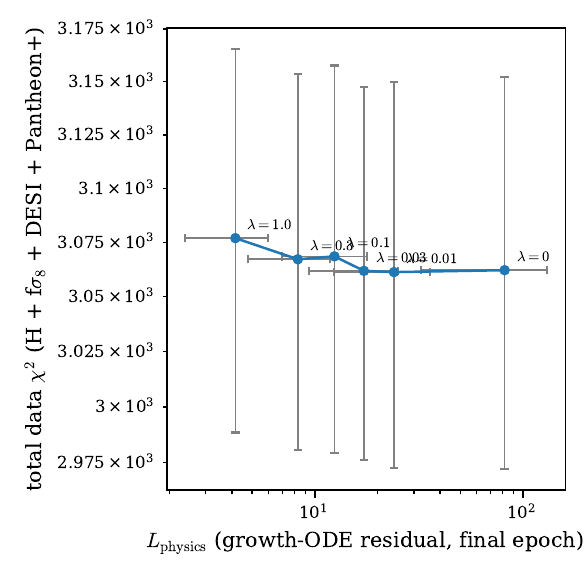}
\caption{L-curve: total data $\chi^2$ (ensemble mean $\pm$ std) versus growth-ODE residual $\mathcal{L}_{\rm physics}$ at convergence, for $\lambda \in \{0,\,0.01,\,0.03,\,0.1,\,0.3,\,1.0\}$. Both axes are logarithmic. The nearly flat $\chi^2$ profile (1.2\% variation across the full $\lambda$ range) indicates that the joint dataset is intrinsically consistent with the growth ODE: enforcing the physics constraint costs almost nothing in data fit. The open circle marks the adopted $\lambda = 0.1$.}
\label{fig:lcurve}
\end{figure}

Table~\ref{tab:ablation} reports the converged ensemble-mean losses and inferred $H_0$ for all six $\lambda$ values. Figure~\ref{fig:lcurve} shows the corresponding L-curve: total data $\chi^2$ versus ODE residual, both averaged over the 100-member ensemble.

\begin{table}[t]
\centering
\caption{Lambda sweep results (no $H_0$ prior, top block) and anchored prior runs (bottom block). $H_0$ and $f\sigma_8(0)$ are ensemble mean $\pm$ standard deviation. Loss values are ensemble means at convergence. The member-to-member standard deviation on total $\chi^2$ is $\approx 89$ at all $\lambda$, larger than the mean shift across the full $\lambda$ range ($+15$).}
\label{tab:ablation}
\begin{tabular}{crrrrr}
\toprule
$\lambda$ &
$H_0$ [km\,s$^{-1}$\,Mpc$^{-1}$] &
$f\sigma_8(0)$ &
Total $\chi^2$ &
$\mathcal{L}_{\rm physics}$ &
$\langle\sigma_H\rangle$\\
\midrule
$0$    & $68.4 \pm 4.8$ & $0.384 \pm 0.114$ & 3062.2 & 81.6 & 23.1 \\
$0.01$ & $67.8 \pm 5.1$ & $0.397 \pm 0.066$ & 3061.3 & 24.0 & 20.7 \\
$0.03$ & $68.3 \pm 4.9$ & $0.405 \pm 0.054$ & 3061.9 & 17.2 & 20.1 \\
$0.1$  & $69.0 \pm 4.7$ & $0.405 \pm 0.047$ & 3068.6 & 12.4 & 19.0 \\
$0.3$  & $68.7 \pm 5.2$ & $0.411 \pm 0.051$ & 3067.2 & \phantom{1}8.3 & 18.7 \\
$1.0$  & $69.5 \pm 4.2$ & $0.404 \pm 0.046$ & 3076.9 & \phantom{1}4.2 & 17.0 \\
\midrule
$0.1$ (SH0ES) & $72.6 \pm 1.0$ & $0.388 \pm 0.035$ & 3065.6 & 12.2 & 22.4 \\
$0.1$ (H0dN)  & $73.2 \pm 0.8$ & $0.383 \pm 0.036$ & 3066.5 & 12.3 & 20.5 \\
\bottomrule
\end{tabular}
\end{table}

Several features stand out. First, $H_0$ is essentially $\lambda$-independent: it ranges only from $67.8$ to $69.5$\,km\,s$^{-1}$\,Mpc$^{-1}$ across all six values. This confirms that $H_0$ is fixed entirely by the data; primarily through the DESI DR2 BAO sound-horizon ratios and Pantheon$+$ distance moduli, rather than by the physics constraint.

Second, the L-curve is nearly flat in total $\chi^2$. Going from $\lambda = 0$ to $\lambda = 1.0$ costs only $+15$ units of $\chi^2$ on a baseline of 3062 (0.5\%), while the ODE residual falls by 95\%. Furthermore, the member-to-member $\chi^2$ spread ($\approx 89$) is larger than the mean shift across all six $\lambda$ values, so the lambdas are statistically indistinguishable by $\chi^2$ alone. This near-flatness is a physically meaningful result: the joint dataset (expansion $+$ growth $+$ BAO $+$ SNe) is intrinsically consistent with the GR linear growth equation, without needing strong enforcement. The physics constraint acts as a lightweight regularizer rather than an informative prior.

Third, the average reconstruction width $\langle\sigma_H\rangle$ (ensemble standard deviation of $H(z)$ averaged over the redshift grid) decreases monotonically with $\lambda$, from 23.1 at $\lambda = 0$ to 17.0 at $\lambda = 1.0$. This improvement is concentrated in data-sparse regions: at $z = 1.2$, for example, the standard deviation falls from $39$\,km\,s$^{-1}$\,Mpc$^{-1}$ ($\lambda = 0$) to $22$\,km\,s$^{-1}$\,Mpc$^{-1}$ ($\lambda = 0.3$), as the ODE provides the interpolation rule between DESI BAO anchor points.

We adopt $\lambda = 0.1$ as the best-value point: it reduces the ODE residual by 85\% relative to the unconstrained case at a $\chi^2$ cost of only $+6$ (0.2\%), and reduces $\langle\sigma_H\rangle$ by 18\%. The physics residual standard deviation across members falls from 60\% of the mean at $\lambda = 0$ (initialization noise dominates) to 44\% at $\lambda = 0.1$, meaning physics enforcement becomes more reproducible across the ensemble.

\subsection{$H(z)$ reconstruction}
\label{sec:hz_result}

\begin{figure}[t]
\centering
\includegraphics[width=\textwidth]{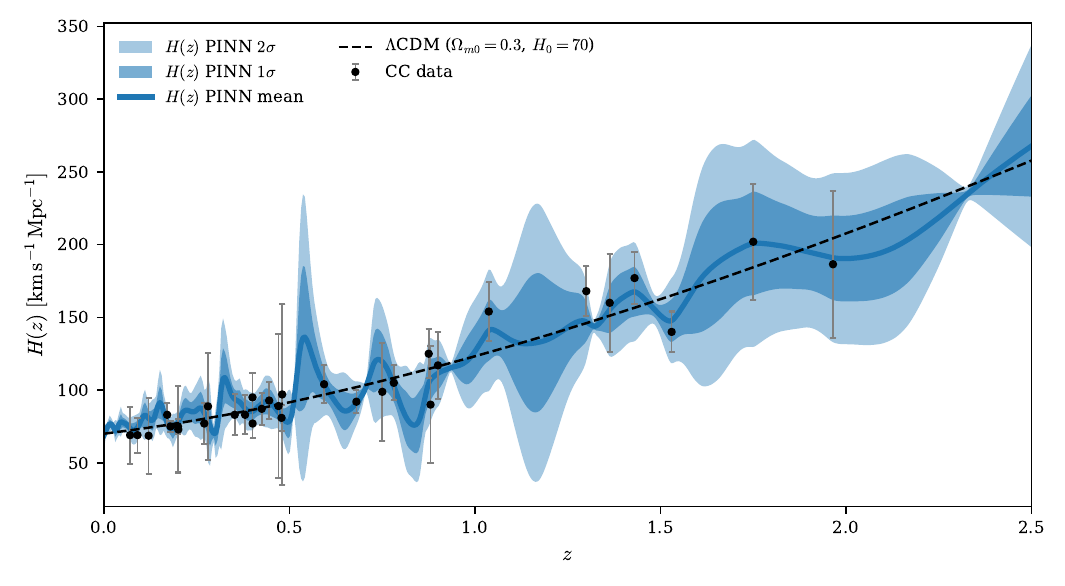}
\caption{$H(z)$ reconstruction (no $H_0$ prior, $\lambda = 0.1$). Darker and lighter bands are $1\sigma$ and $2\sigma$ ensemble spread from 100 members. The dotted curve is the Planck 2018 $\Lambda$CDM prediction. Points show the 32 CC measurements.}
\label{fig:hz_noPrior}
\end{figure}

Figure~\ref{fig:hz_noPrior} shows the $H(z)$ reconstruction for the no-prior run at $\lambda = 0.1$, overlaid on the CC data and the flat $\Lambda$CDM prediction for $\Omega_{m,0} = 0.315$, $H_0 = 67.4$\,km\,s$^{-1}$\,Mpc$^{-1}$ (Planck 2018).
The free reconstruction yields $H_0 = 69.0 \pm 4.7$\,km\,s$^{-1}$\,Mpc$^{-1}$ (ensemble median $70.7$, 16th--84th percentile range $64.0$--$73.4$). This is consistent with the Planck 2018 CMB value ($67.4 \pm 0.5$) at well under $1\sigma$, and approximately $0.9\sigma$ below the SH0ES local measurement ($73.04 \pm 1.04$). This constitutes an inverse distance-ladder determination of $H_0$: the absolute scale is set by the DESI DR2 BAO sound-horizon ratios and the Pantheon$+$ luminosity distances, with no Cepheid input.

The shape of $H(z)$ shows a systematic depression relative to $\Lambda$CDM at $z \sim 0.5$--$0.8$. At $z = 0.8$ the ensemble mean lies approximately $18$\% below the $\Lambda$CDM curve, at $\approx 1.4\sigma$ significance. This deficit is  robust: it is present at every $\lambda$ value in Table~\ref{tab:ablation} (ranging from $-1.4\sigma$ at $\lambda = 0.1$ to $-1.8\sigma$ at $\lambda = 1.0$), and therefore cannot be attributed to the physics enforcement. It is driven by the DESI DR2 LRG2 $D_H/r_s$ constraint at $z_{\rm eff} = 0.706$ ($D_H = c/H$, so a large $D_H$ implies a low $H$), and is qualitatively consistent with the dark energy evolution reported by DESI DR2 when fitting $w_0w_a$CDM \cite{DESI:2025zgx}.

\begin{figure}[t]
\centering
\includegraphics[width=\textwidth]{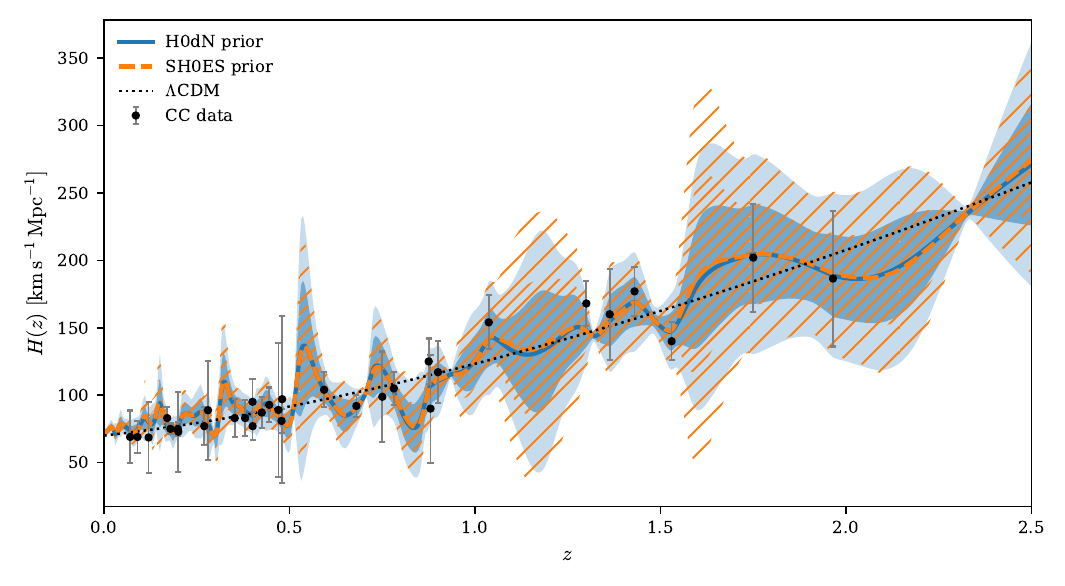}
\caption{$H(z)$ reconstruction for the H0dN prior
($H_0 = 73.50 \pm 0.81$\,km\,s$^{-1}$\,Mpc$^{-1}$, blue filled bands) and the SH0ES prior ($H_0 = 73.04 \pm 1.04$\,km\,s$^{-1}$\,Mpc$^{-1}$, orange hatched bands), both at $\lambda = 0.1$. Darker and lighter bands are $1\sigma$ and $2\sigma$. The two anchored reconstructions are nearly indistinguishable in $H(z)$ shape; the difference is confined to the $H_0$ value enforced at $z = 0$.}
\label{fig:hz_comparison}
\end{figure}

At $z \gtrsim 1$ the reconstruction is consistent with $\Lambda$CDM within $1\sigma$. At $z = 2.3$--$2.5$, the Ly$\alpha$ BAO anchor provides a loose upper constraint; the bands widen as the distance integral accumulates uncertainty. Figure~\ref{fig:hz_comparison} compares the no-prior reconstruction with the two
anchored runs.

\subsection{$f\sigma_8(z)$ reconstruction}
\label{sec:fs8_result}

\begin{figure}[t]
\centering
\includegraphics[width=\textwidth]{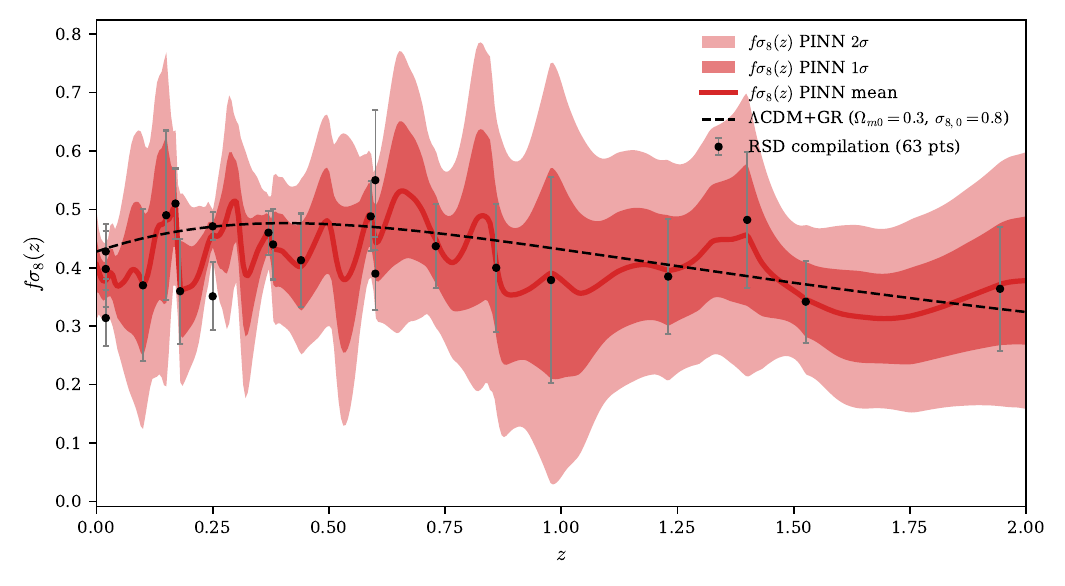}
\caption{$f\sigma_8(z)$ reconstruction (no $H_0$ prior, $\lambda = 0.1$). Bands are $1\sigma$ and $2\sigma$ ensemble spread. The dotted curve is the $\Lambda$CDM$+$GR prediction integrated with $\Omega_{m,0} = 0.315$, $\sigma_{8,0} = 0.8111$ (Planck 2018), peaking at $z \approx 0.5$ due to the competition between rising growth rate $f(z)$ and falling growth factor $D(z)$. Points show the 22 RSD measurements.}
\label{fig:fs8_noPrior}
\end{figure}

Figure~\ref{fig:fs8_noPrior} shows the $f\sigma_8(z)$ reconstruction for the no-prior run at $\lambda = 0.1$.
The reconstruction is broadly consistent with the $\Lambda$CDM$+$GR prediction in the data-dense range $z \sim 0.3$--$0.6$ ($f\sigma_8 \approx 0.44$--$0.51$, ensemble $1\sigma \approx \pm 0.07$--$0.09$). At $z = 0$ the reconstruction extrapolates to $f\sigma_8(0) = 0.405 \pm 0.047$; since no RSD measurements exist at $z = 0$, this value is purely network-driven. At $z \gtrsim 1.2$ the uncertainties widen considerably as the RSD data become sparse.

\begin{figure}[t]
\centering
\includegraphics[width=\textwidth]{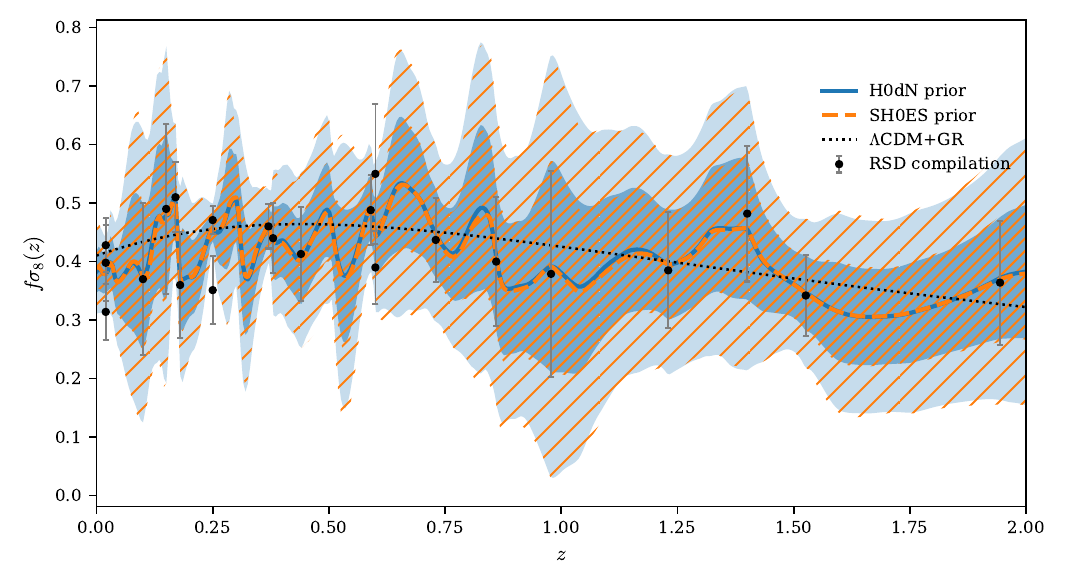}
\caption{$f\sigma_8(z)$ reconstruction for the H0dN prior (blue filled bands) and the SH0ES prior (orange hatched bands), both at $\lambda = 0.1$. Darker and lighter bands are $1\sigma$ and $2\sigma$. Both anchored runs yield a mild suppression of $f\sigma_8(0)$ relative to the free reconstruction ($\approx 0.385$ vs $0.405$), with the two prior runs essentially overlapping across the full redshift range.}
\label{fig:fs8_comparison}
\end{figure}

Figure~\ref{fig:fs8_comparison} overlays the two anchored conditional reconstructions. With $H_0$ anchored to $\approx 73$\,km\,s$^{-1}$\,Mpc$^{-1}$, the reconstructed $f\sigma_8(0)$ is mildly suppressed relative to the free run (from $0.405$ to $\approx 0.385$, a $\sim 0.5\sigma$ shift) through the growth ODE: a higher $H_0$ modifies the factor $\Omega_m(a)$ in Eq.~\eqref{eq:growth_ode}, and the network adjusts $f\sigma_8$ at low $z$ accordingly. This illustrates the $H_0$--$\sigma_8$ link propagated through the ODE coupling, though the effect is well within the current uncertainty bands.

\subsection{Ensemble quality}
\label{sec:spread}

\begin{figure}[t]
\centering
\includegraphics[width=0.75\textwidth]{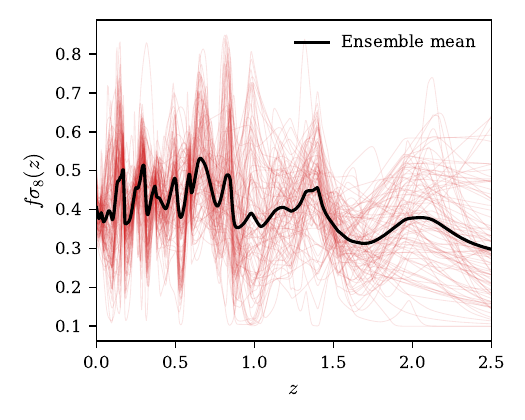}
\caption{Individual $f\sigma_8(z)$ trajectories for all 100 ensemble members (thin lines, no $H_0$ prior, $\lambda = 0.1$), overlaid with the ensemble mean (thick black).}
\label{fig:spread}
\end{figure}

Figure~\ref{fig:spread} shows all 100 individual $f\sigma_8(z)$ trajectories for the no-prior $\lambda = 0.1$ run. The ensemble is well-behaved: all members follow a broadly similar shape with no evidence of multimodality. The spread is well-controlled below $z \sim 1.5$ and widens at higher redshifts where the RSD data are sparse. The bootstrap resampling and fiducial-cosmology draws produce a range of member trajectories that is visibly larger than in a seed-only ensemble, reflecting the more complete uncertainty budget.

\section{Discussion and Conclusions}
\label{sec:discussion}

The free reconstruction yields $H_0 = 69.0 \pm 4.7$\,km\,s$^{-1}$\,Mpc$^{-1}$, fully consistent with the Planck 2018 CMB value at well under $1\sigma$. This is not a model-fit to CMB data: it is derived entirely from late-universe observables (DESI DR2 BAO, Pantheon$+$ supernovae, CC chronometers, and RSD growth rates) through a model-independent network reconstruction. The BAO sound-horizon ratios and supernova luminosity distances jointly anchor the absolute distance scale, recovering a Planck-consistent $H_0$ without any Cepheid input. The result is approximately $0.9\sigma$ below SH0ES ($73.04 \pm 1.04$), consistent with the accumulating evidence that the inverse distance ladder and the Planck CMB are mutually consistent, and that the Hubble tension lives in the Cepheid-calibration rung of the forward distance ladder. The result is conditional on the Planck 2018 sound horizon $r_d = 147.09 \pm 0.26$\,Mpc used to calibrate the BAO distance ratios; $H_0$ scales as $c/r_d$, so widening to a CMB-independent BBN calibration ($r_d$ uncertainty $\sim 3$\,Mpc) would broaden the $H_0$ posterior to $\pm 8$--$10$\,km\,s$^{-1}$\,Mpc$^{-1}$ while shifting the center by $\lesssim 2$\,km\,s$^{-1}$\,Mpc$^{-1}$.

A notable feature in the reconstruction is the $\approx 1.6\sigma$ depression of $H(z)$ relative to flat $\Lambda$CDM at $z \sim 0.7$--$0.8$, with the ensemble mean lying $\approx 18$\% below the $\Lambda$CDM curve at $z = 0.8$. This is driven by the DESI DR2 LRG2 $D_H/r_s$ measurement at $z_{\rm eff} = 0.706$, which constrains $H(z)$ directly via $D_H = c/H$. The feature is robust across all six $\lambda$ values (Table~\ref{tab:ablation}) and therefore cannot be attributed to the physics-enforcement scheme. It is qualitatively consistent with the departure from $\Lambda$CDM reported by DESI DR2 in a $w_0w_a$CDM fit \cite{DESI:2025zgx}, though at current precision the deviation does not reach the $2\sigma$ threshold. The physical interpretation, i.e. whether this reflects dynamical dark energy, a modified expansion at intermediate redshifts, or residual systematic uncertainties in the BAO measurements, requires a dedicated model-comparison analysis, which we defer to future work.

The two anchored runs (SH0ES and H0dN priors) demonstrate the propagation of the Hubble tension into the growth sector through the ODE coupling. Anchoring $H_0 \approx 73$ mildly suppresses $f\sigma_8(0)$ relative to the free reconstruction (from $0.405$ to $\approx 0.385$, a $\sim 0.5\sigma$ effect) through  Eq.~\eqref{eq:growth_ode}, with the SH0ES and H0dN runs producing essentially the same growth history at $z > 0.1$. These runs are conditional analyses, meaning they show what the growth history must look like \textit{if} the SH0ES $H_0$ is correct, not independent constraints. The primary result remains the free reconstruction.

The near-flat L-curve (0.5\% total $\chi^2$ variation across $\lambda \in [0,\,1]$) is itself a physically informative result. It demonstrates that the joint dataset is intrinsically consistent with the GR linear growth equation: the network satisfies the ODE approximately even without explicit enforcement, and the physics constraint acts as a lightweight regularizer rather than an informative prior. This directly addresses the concern that the physics-loss weight might be doing uncharacterized work: the L-curve shows precisely what work it is doing, and how much data-fit it costs. 

The ensemble spread captures three sources simultaneously: network-initialization variance, data-noise uncertainty (via bootstrap resampling), and fiducial-cosmology systematics (via per-member Planck-prior draws of  $\Omega_{m,0}$, $\sigma_{8,0}$, $r_d$). The current bands are a more honest representation of what the data can constrain model-independently than a seed-only ensemble.

The reconstruction assumes flat geometry and the GR linear growth equation. The fiducial parameters $\Omega_{m,0}$, $\sigma_{8,0}$, and $r_d$ are varied across ensemble members using Planck 2018 priors; their collective impact on the reconstruction is included in the quoted uncertainties. The term ``model-independent'' is used in the restricted sense of assuming no dark energy equation of state and no parametric form for $H(z)$ or $f\sigma_8(z)$.

The current $f\sigma_8$ compilation \cite{Akarsu:2025ijk} already incorporates post-2018 eBOSS DR16 results at $z \sim 0.7$--$1.5$ \cite{eBOSS:2020yzd}, sharpening the reconstruction in exactly the redshift range where the $H(z)$ tension is most visible. The most important near-term extension is therefore to add DESI DR2 RSD measurements of the growth rate \cite{DESI:2025zgx} once they become available, which would provide a direct cross-check of the BAO-driven $H(z)$ depression with an independent growth-sector observable at the same redshifts. Whether any coherent departures from $\Lambda$CDM are consistent with a sign-switching dark energy density \cite{Akarsu:2021fol} is a natural model-comparison question for a follow-up analysis.

\acknowledgments

Project BridgingCosmology is financed by Xjenza Malta and the Scientific and Technological Research Council of T\"{U}B\.{I}TAK, through the Xjenza Malta--T\"{U}B\.{I}TAK 2024 Joint Call for R\&I projects. This article is based upon work from COST Action CA21136 \emph{Addressing observational tensions in cosmology with systematics and fundamental physics} (CosmoVerse), supported by COST (European Cooperation in Science and Technology).

\appendix

\section{Null tests}
\label{app:nulltests}

We collect here two null tests of the reconstruction against the flat   $\Lambda$CDM expectation. Neither test enters the training loop; both are computed after the fact from the converged no-prior, $\lambda = 0.1$ ensemble. The $\mathrm{Om}(z)$ test probes the expansion history alone, while the $\mathrm{Om}_{f\sigma_8}(z)$ test probes the growth sector, so together they check the two reconstructed functions along independent lines.

\subsection{$\mathrm{Om}(z)$ null test}
\label{app:Om}

\begin{figure}[t]
\centering
\includegraphics[width=\textwidth]{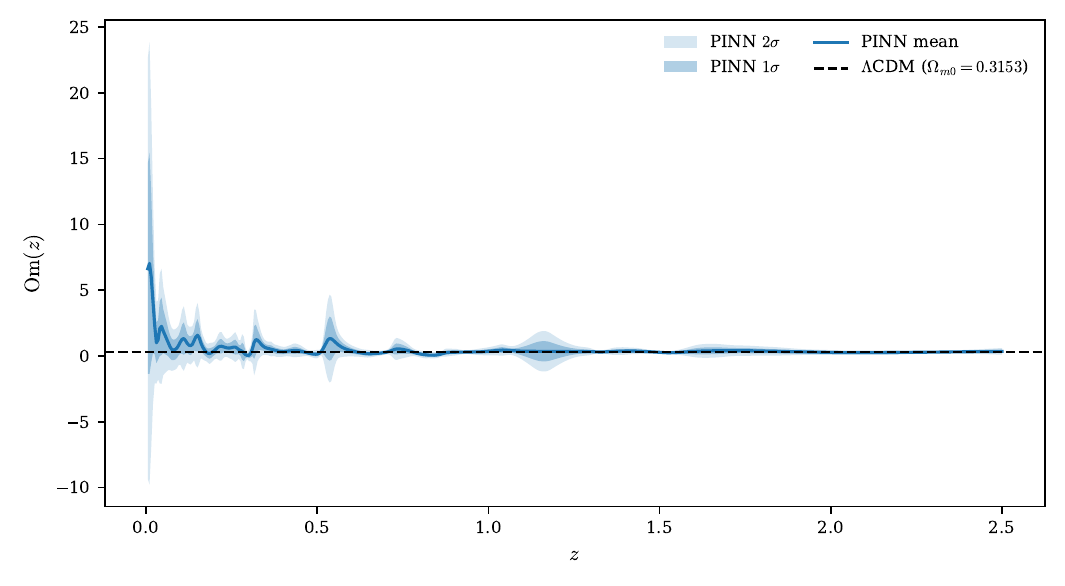}
\caption{$\mathrm{Om}(z)$ null test for the no-prior reconstruction ($\lambda = 0.1$). Darker and lighter bands are $1\sigma$ and $2\sigma$ ensemble spread. The dashed horizontal line marks the Planck 2018 value $\Omega_{m,0} = 0.315$.}
\label{fig:Om}
\end{figure}

The $\mathrm{Om}(z)$ diagnostic \cite{Sahni:2008xx} provides a purely kinematic test of the expansion history:
\begin{equation}
    \mathrm{Om}(z) = \frac{E^2(z) - 1}{(1+z)^3 - 1}\,,\qquad E(z) = \frac{H(z)}{H_0}\,,
    \label{eq:Om}
\end{equation}
where $H_0 = H(z=0)$ is taken from the reconstruction. For flat $\Lambda$CDM, $\mathrm{Om}(z) = \Omega_{m,0}$ exactly; any redshift dependence signals dynamical dark energy or a departure from GR. A profile rising with $z$ indicates $w > -1$; falling indicates $w < -1$. The diagnostic is computed entirely from the reconstructed $H(z)$ profile and is insensitive to the $\Omega_{m,0}$ prior, which enters only through the weak ODE coupling to the $f\sigma_8$ head. Figure~\ref{fig:Om} shows $\mathrm{Om}(z)$ for the no-prior $\lambda = 0.1$ reconstruction.

\subsection{$\mathrm{Om}_{f\sigma_8}(z)$ null test}
\label{app:Omfs8}

\begin{figure}[t]
\centering
\includegraphics[width=\textwidth]{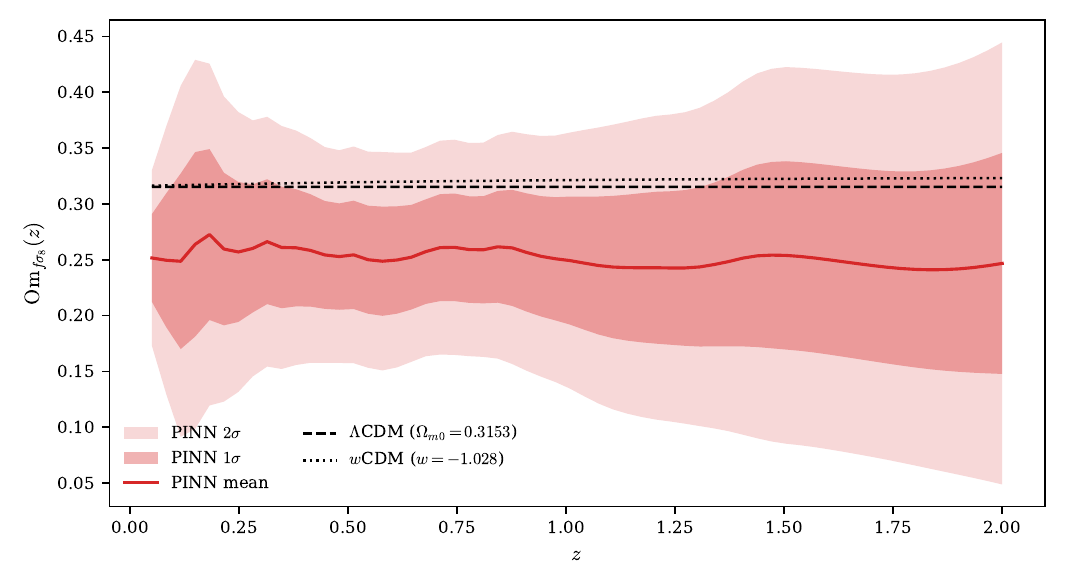}
\caption{$\mathrm{Om}_{f\sigma_8}(z)$ null test for the no-prior reconstruction ($\lambda = 0.1$). Bands are $1\sigma$ and $2\sigma$. The dashed curve is the $\Lambda$CDM expectation for $\Omega_{m,0} = 0.315$.}
\label{fig:Omfs8}
\end{figure}

The $\mathrm{Om}_{f\sigma_8}(z)$ diagnostic \cite{Dialektopoulos:2021wde} tests the growth sector independently of the expansion history. From the reconstructed $f\sigma_8(z)$ one integrates 
\begin{equation}
  \frac{{\rm d}\Delta_m}{{\rm d}\ln a} = \frac{f\sigma_8(z)}{\sigma_{8,0}}\,,
  \qquad \Delta_m(0) = 1\,,
\end{equation}
to recover the normalized growth factor $\Delta_m(z)$, then finds the effective $\Omega_{m,0}$ that makes the flat $\Lambda$CDM growth factor match $\Delta_m(z)$ at each redshift by root-finding. The resulting profile should be flat and equal to the true $\Omega_{m,0}$ under $\Lambda$CDM$+$GR. Note that this diagnostic is sensitive to the $\Omega_{m,0}$ prior through the ODE coupling to the $f\sigma_8$ head, and its output should be interpreted as a test of internal consistency given the Planck-prior fiducial cosmology rather than an independent measurement of $\Omega_{m,0}$. Figure~\ref{fig:Omfs8} shows $\mathrm{Om}_{f\sigma_8}(z)$ for the no-prior run.

\bibliographystyle{unsrtnat}
\bibliography{paper}

\end{document}